# Retrieval-Augmented Generation in LLMs for Mental Health: Quantifying the Incremental Contribution of Retrieval Within a Layered Safety Architecture

Anand Gupta[a], Akshat Surolia[a], Shubham Mishra[a], Shakil Imtiaz[a], & Chaitali Sinha[a]
[a] *Wysa Inc., Boston, MA, United States*

## Abstract

Digital mental health interventions (DMHIs) offer scalable support, but ensuring they accurately detect users' intent during volatile situations can be challenging. Pure parametric Large Language models (LLMs) do not contain specific safety critical architecture, and can miss critical cues, or hallucinate, undermining reliability. Retrieval Augmented Generation (RAG), which supplements an LLM with retrieved context, could enhance intent detection during volatile situations. Commercially available DMHIs typically combine multiple independent safety layers like rule-based filters, symbolic escalation protocols, and neural classification. The incremental contribution of any single layer, however, remains unquantified. This paper evaluates six LLM models within a DMHI called Wysa, via a controlled comparison of RAG-enabled versus RAG-disabled modes. Anonymized real and synthetic user-chatbot exchanges were annotated by a qualified clinical team against multi-class intent categories (e.g. self-harm, abuse, panic). The study computed classification accuracy, recall, precision and F1 scores against ground truth labels and tested differences for statistical significance. Performance was also examined by risk category and inter-model agreement. While RAG caused a rise in false alarms, the trade-off is consistent with safety-critical design principles that prioritize sensitivity, where flagged cases are routed to additional review rather than acted on directly. Overall, these findings support RAG as a promising approach to improve the accuracy, consistency and safety of LLM-driven DMHIs.

## Background

Mental disorders are one of the leading contributors to the global health burden, with approximately one-third of people experiencing poor health in their lifetime.[1] This growing demand for mental health services and limited access to qualified care providers require the search for easily available, accessible resources and alternate modalities of care.[2]

In this context, the present age of digitalization brings progress and new possibilities for mental health and psychotherapy.[3] The use of artificial intelligence in psychotherapy offers the potential to simulate human-like care and nuance, especially with intelligent therapy chatbots that adapt to individual user-needs and provide continued support. Such digital mental health interventions (DMHIs) including internet and smartphone-delivered services, hold promise for overcoming significant barriers that are traditionally associated with face-to-face mental health care,[4,5] such as stigma,[6] accessibility and cost.[7]

Within this landscape, large language models (LLMs) now underpin many DMHI solutions, helping approximate human-level performance on certain tasks to understand and generate language. However, their purely parametric nature, i.e. relying solely on knowledge encoded within model weights, poses significant challenges for clinical reliability. Furthermore, their propensity to hallucinate can create serious repercussions within mental health settings.

### Retrieval Augmented Generation

In the middle of 2020, Lewis et al[8] introduced Retrieval Augmented Generation (RAG) constituting a significant advancement in LLM architectures, addressing generative task limitations through retrieval mechanisms. Within mental health applications, RAG demonstrates particular efficacy in mitigating parametric model deficiencies, namely hallucination propensity and contextual inadequacy.

The intervention employs a layered, neurosymbolic safety architecture in which retrieval-augmented classification operates alongside independent rule-based filters, deterministic escalation protocols, and human-in-the-loop review. No single layer is solely responsible for safety outcomes. The present study isolates the retrieval layer for controlled evaluation; all reported figures reflect this component in isolation and do not represent end-to-end system performance. Within this layered architecture, the retrieval component encompasses two primary operational domains: contextual continuity preservation and clinical safety classification.

For contextual continuity, the system employs conversational history spanning preceding $n^{th}$ user interactions, systematically retrieving relevant context to enhance semantic understanding of discourse trajectory. This is essential for chatbot services which may have to effectively distinguish between cases of severe symptomatology, or an instance of self-disclosure. By leveraging context from earlier conversations, RAG ensures that such situations are mitigated through appropriate structured, evidence-based routes.

Regarding clinical safety, the system utilizes LLM-based classification whereby conversational context undergoes systematic evaluation for explicit or implicit risk indicators, subsequently initiating appropriate crisis intervention protocols while maintaining adherence to non-diagnostic operational parameters. This targeted RAG methodology significantly improves language model accuracy on clinical tasks, effectively addressing core limitations like hallucination tendencies. Similar comparisons by Béchard and Ayala[9] report that incorporating RAG into a generative workflow system significantly reduces hallucination and even allows using a smaller LLM without performance loss.[10]

While generative models show anecdotal promise in producing empathetic, evidence-based responses, there is a lack of formal evaluation in interactive mental health contexts, particularly comparing RAG systems with standalone publicly available LLMs. Existing literature from adjacent domains suggests RAG offers advantages when domain alignment is critical, but this has not been empirically validated in mental health support scenarios. This study presents a controlled ablation of the retrieval layer within a commercially deployed DMHI's multi-layered safety architecture, quantifying retrieval's incremental contribution to intent classification across six LLMs of varying capacity.

## Methodology

### About the Intervention

The evaluated intervention is a commercially deployed, evidence-based conversational agent delivering CBT, behavioral reinforcement, and mindfulness techniques through a combination of rule-based and LLM components. Its safety architecture comprises multiple independent layers; the present evaluation isolates the retrieval-augmentation component.

### Data Collection and Processing

The study utilizes two primary data sources, (1) a corpus of PII-redacted (anonymized) real user-chatbot conversations capturing authentic user interactions across a range of clinical risk presentations; and (2) a curated set of synthetic text prompts and dialogue segments derived from verified real user risk conversations. The synthetic corpus was generated by applying a structured re-synthesis of Wysa's existing repository of anonymized real user interactions spanning both pre-labelled true and false distress. This methodology ensures that the synthetic data preserves the natural linguistic and contextual characteristics of genuine user expression, including colloquial phrasing, indirect disclosures, and ambiguous risk language, whilst precluding any re-identification of individuals.

All conversation data was refactored and standardized before analysis. All logs were curated into an ordered list with alternating user and bot messages for consistent, contextualized processing and finally stored in a JSON format. The metadata was filtered to only retain conversational segments wherein intent was identifiable (i.e. discarding entries with single or incomplete words that do not make literal sense). This filtering ensured that all evaluated dialogues included at least one prompt that triggered the retrieval mechanism. Finally, the dataset was checked for semantic de-duplication to remove duplicate or nearly identical query instances, so that evaluation is not biased by repeated content. The resulting cleaned set was then used as the input for the evaluation paradigms described below.

## Evaluation Metrics

The evaluation encompassed a controlled classification task for moderation and safety intent detection. This design constitutes a single-layer ablation. All other safety components of the production system, including rule-based filters and escalation protocols were excluded from this evaluation to isolate the retrieval layer's marginal contribution. The accuracy and recall figures reported throughout therefore, represent the retrieval component's individual contribution and must not be interpreted as the end-to-end detection performance of the production safety system.

This was conducted using an open-ended response comparison wherein the entire dataset was divided into two parts, the minority RAG collection using a subset for retrieval and the test set (used interchangeably as user query below). In practice, the user query was presented to the chatbot twice - first in the RAG-augmented mode, and then with the RAG system turned off (i.e. using only the base LLM without retrieval augmentation). By rerunning the exact chat segments in both modes, a paired bot response was obtained for direct comparison.

The test evaluation (including user query) focused on multi-class classification critical to a mental health chatbot's safety. Several thousand individual user utterances covering a broad range of moderation-sensitive and safety-critical scenarios, were coded. Each user log was annotated with a ground-truth label called "Risk category." These labels were established through a systematic annotation by Wysa's clinical team of qualified therapists, with each entry independently assigned a risk category in accordance with the taxonomy defined in Table 1.

| Risk Category | Definition |
|---|---|
| Illegal activity | Mentions or endorsements of actions that are against the law. |
| Harm towards others | Statements indicating intent or risk of physical, emotional, or verbal harm directed at other individuals. |
| Abuse towards child | Mentions of physical, emotional, or sexual harm inflicted on a minor. |
| Panic attack | Language suggestive of an acute episode of intense fear or physiological distress. |
| Trauma | Disclosures or references to distressing past experiences likely to be psychologically triggering. |
| Prompt injection | Attempts to manipulate or override the chatbot's behavior through adversarial or meta-level instructions. |
| No risk | Content that poses no identifiable psychological, safety, or policy concerns. |

**Table 1:** Multi-class Risk Taxonomy Used for Intent Classification and Safety Labeling

## Analysis

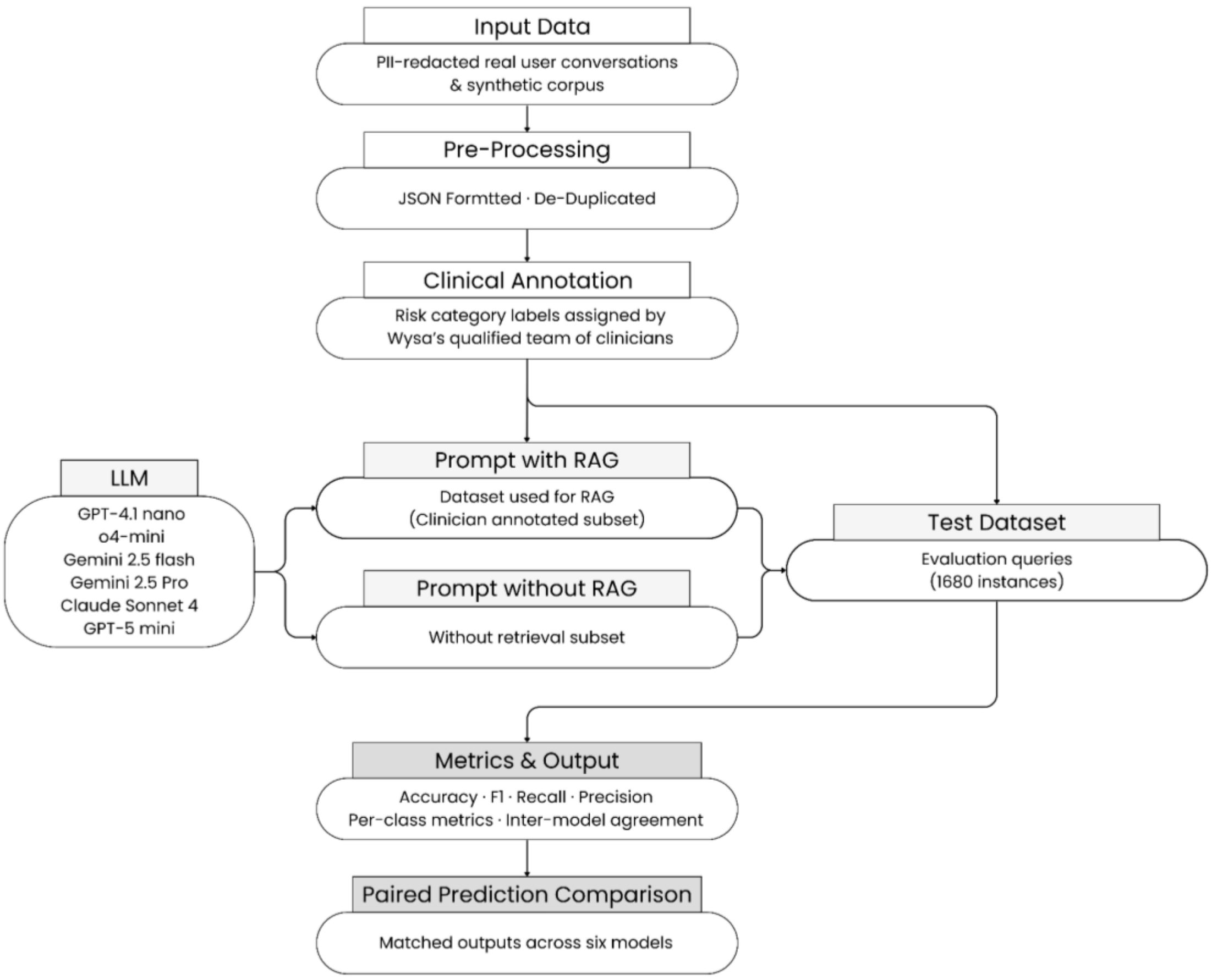

**Figure 1:** Study Flow

Using these two configurations (RAG versus No-RAG) under a deterministic prompt (i.e. temperature = 0), a reproducible set of results was obtained. In the RAG condition, the system could retrieve supplemental context (from the clinician-annotated subset used to build the RAG index) before producing a label, whereas the tests without RAG relied solely on the model's internal knowledge. The performance for each configuration was measured using standard metrics like classification accuracy, F1 score, and Cohen's Kappa[11], comparing the model-predicted labels against ground truth annotations established by qualified clinical raters. To assess significance of any performance differences intra-cohort (i.e. under the same LLM), McNemar's test[12] on paired outcomes was computed. Confidence intervals[13] were calculated via bootstrap resampling, providing an estimate of uncertainty around the measured performance. Additionally, confusion matrices for each condition were analyzed to identify any shifts in error patterns when RAG was enabled (e.g. to see if retrieval context reduced false negatives on critical safety alerts), and delta tests were used to directly quantify the net contribution of retrieval augmentation by comparing RAG and No-RAG conditions. The results were examined across the different LLMs (inter-cohort) to evaluate the consistency of RAG's impact, helping determine observed performance gains (or lack thereof) across model architectures or specific to a singular model variant. Finally, to assess per-class performance, 95% Wilson confidence intervals were computed for each class-level recall estimate, using the true positive count and class support as inputs. Per-class McNemar's tests were additionally applied to evaluate the statistical significance of recall differences between RAG and No-RAG conditions for each risk category independently, with significance thresholds set at $p < 0.05$, $p < 0.01$, and $p < 0.001$. The complete analytical workflow is illustrated in Figure 1.

Since the evaluation produced paired, discrete, multi-class predictions against a fixed ground truth, the choice of statistical methods was shaped accordingly. Classification accuracy, F1 score, and Cohen's Kappa were used for overall performance: accuracy and F1 capture raw and class-weighted correctness respectively, while Kappa additionally corrects for chance agreement, which is essential given the imbalanced label distribution where the dominant "No Risk" class would otherwise inflate apparent consensus. McNemar's test was applied for significance testing as it is specifically designed for paired binary outcomes, directly appropriate for a within-model, within-input RAG versus No-RAG comparison. Bootstrap resampling was used for confidence interval estimation given its assumption-light properties under non-normal classification data, and Wilson confidence intervals were selected for per-class recall estimation due to their superior coverage properties at boundary proportions.

## Results

Across the six language models tested, RAG boosted the overall classification performance and increased accuracy for 5 of 6 models, with large gains especially for the lightweight OpenAI models. For instance, with all other safety layers excluded, GPT-4.1 nano's accuracy jumped from 48.3% to 72.7% with RAG (Δ +24.5 percentage increase, with a 95% Confidence Interval of 21.7 - 27.2%). This improvement is highly significant with a McNemar test value of $\chi^2 \approx 256$, $p \ll 0.001$. o4-mini also saw a significant accuracy increase with a bump from 71.4% to 79.2% (Δ +7.8 percentage increase, with a 95% CI of 5.7 - 9.8%).

| X (No RAG) | Y (RAG) | A | B | C | D | Discordant | B/C | NDR | Acc_X | Acc_Y | $\chi^2$ |
|---|---|---|---|---|---|---|---|---|---|---|---|
| GPT-4.1 nano | GPT-4.1 nano | 687 | 124 | 535 | 334 | 659 | 0.232 | -0.244 | 0.482 | 0.727 | 256.329 |
| o4-mini | o4-mini | 1118 | 81 | 212 | 269 | 293 | 0.382 | -0.077 | 0.713 | 0.791 | 58.569 |
| Gemini 2.5 flash | Gemini 2.5 flash | 1224 | 88 | 154 | 214 | 242 | 0.571 | -0.039 | 0.781 | 0.820 | 18 |
| Gemini 2.5 Pro | Gemini 2.5 Pro | 1121 | 189 | 172 | 198 | 361 | 1.099 | 0.0101 | 0.779 | 0.769 | 0.801 |
| Claude Sonnet 4 | Claude Sonnet 4 | 1203 | 76 | 184 | 217 | 260 | 0.413 | -0.064 | 0.761 | 0.825 | 44.861 |
| GPT-5 mini | GPT-5 mini | 1231 | 87 | 114 | 248 | 201 | 0.763 | -0.016 | 0.784 | 0.801 | 3.626 |

**Table 2** McNemar's Test comparing paired classification outcomes with and without RAG across models

Where,
**A:** Both models correct
**B:** RAG - Incorrect, No RAG - Correct
**C:** RAG - Correct, No RAG - Incorrect
**D:** Both models incorrect
**NDR (Net Disagreement Rate)**: (B - C) / N
**Acc_X:** No RAG Accuracy, (A + B) ÷ N
**Acc_Y:** RAG Accuracy, (A + C) ÷ N
**McNemar $\chi^2$:** $(B - C)^2 \div (B + C)$

The larger Claude Sonnet 4 showed positive but smaller gains, with accuracy changing from 76.13% to 82.56% (Δ +6.43%, $p \ll 0.001$, with a 95% CI of 4.5% - 8.2%). Both Gemini 2.5 Pro and GPT-5 mini however, did not show statistically significant changes. Overall, RAG consistently and significantly improved accuracy for four models (GPT-4.1 nano, o4-mini, Gemini 2.5 flash, Claude Sonnet 4), had a negligible impact on one (Gemini 2.5 Pro) and yielded a minor, non-significant uptick for the most advanced model (GPT-5 mini).

Similar trends were also mirrored in the precision, recall and $F_1$ scores. Globally RAG tended to improve recall more substantially than precision, leading to a higher $F_1$ in most cases. It was observed that RAG's benefits were inversely correlated with baseline performance, such that lightweight models gained the most, and advanced models gained the least. The only model with essentially unchanged $F_1$ was Gemini 2.5 Pro (macro $F_1 \sim 0.79$ in both conditions), consistent with its flat accuracy Δ.

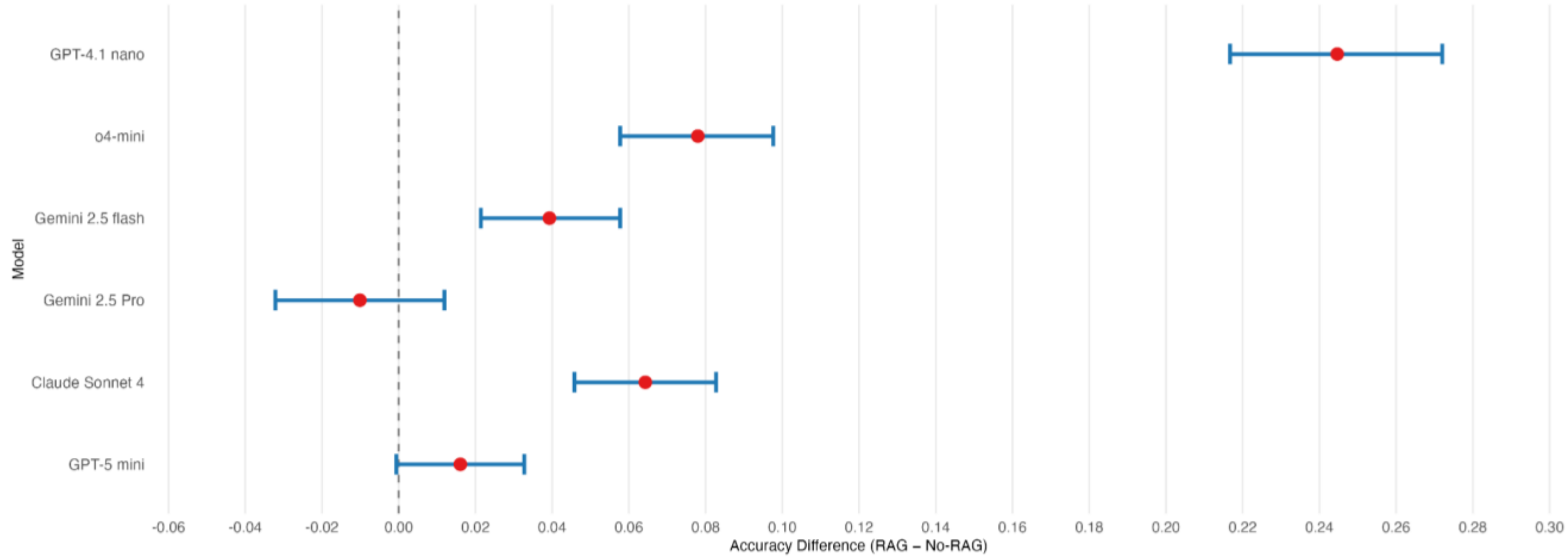

**Figure 2:** Bootstrap 95% confidence intervals (CI) for accuracy gain (Δ = RAG − No-RAG) across six models. *Values above zero indicate improvement with RAG; intervals crossing zero indicate non-significant change.*

These global patterns indicate that RAG provided a net performance boost in most settings, chiefly by helping models retrieve and correctly classify examples they initially missed (improved recall) with only minor trade-offs in precision.

## Class-Level Results

RAG especially improved performance with classes that models originally struggled with reducing instances of under-detection. For example, with retrieval disabled and no other safety layers active, GPT-4.1 nano correctly recognized only 2.7% of instances of “Abuse towards child” in the test set, and 16.3% of “Panic attack” in cases without RAG, indicating that it was missing the vast majority of those at-risk queries. These patterns are evident in the confusion matrix, where many actual cases were being misclassified into the benign “No risk” category (explaining the near 87% recall for No risk). However, with RAG, the model’s sensitivity to these critical classes dramatically improved (“Abuse towards child” from 2.7% to 50.0%, and “Panic attack” from 16.3% to 73.6%). This represents a 47–57-point increase in recall for those cohorts. Correspondingly, the $F_1$ for “Panic attack” soared from a low 0.28 to 0.78 with RAG. Other analogous improvements are also showcased in other models, leading to the conclusion that RAG consistently helped models catch more positive instances of volatile intent, thereby reducing false-negatives in those classes.

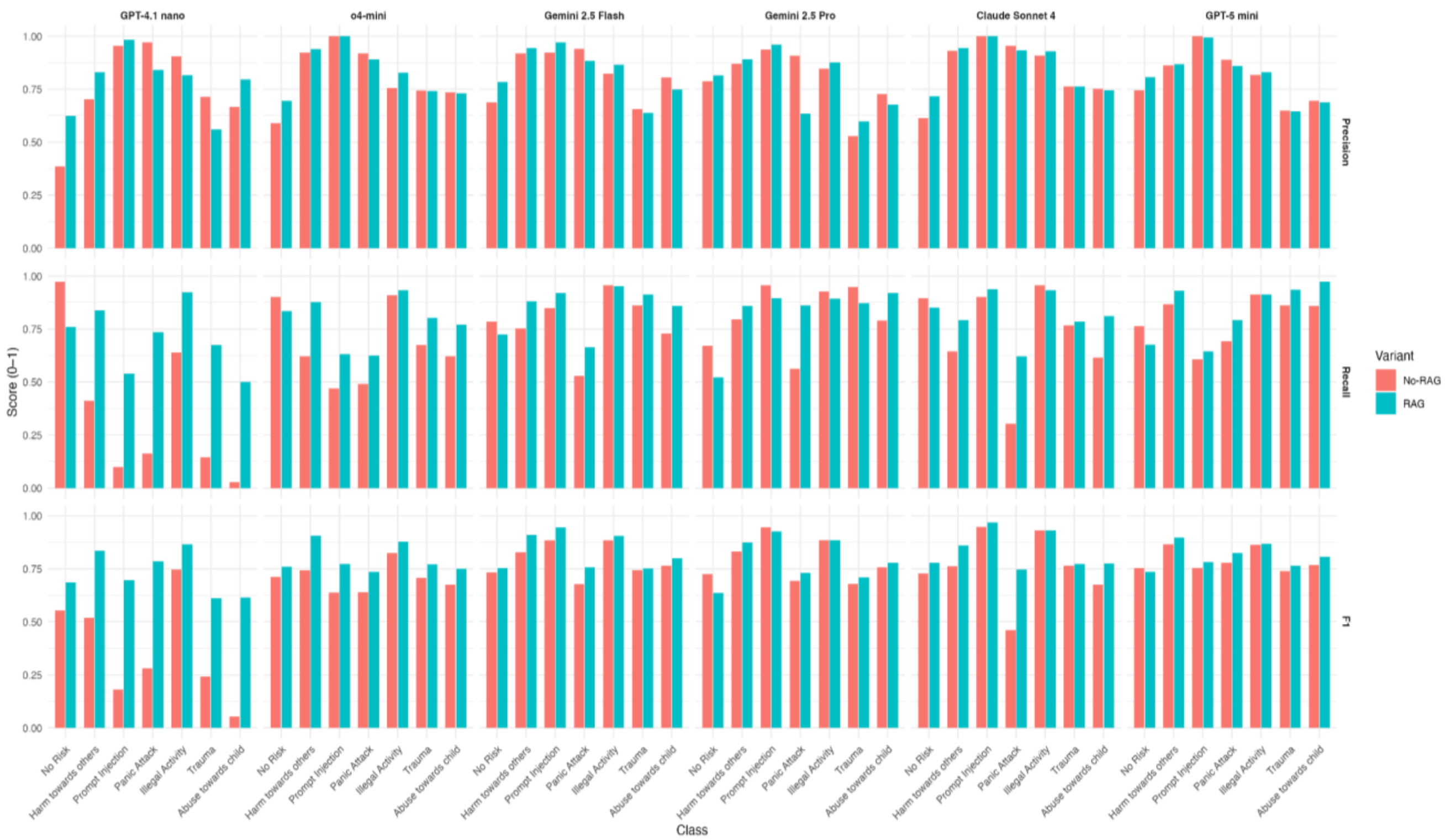

**Figure 3:** Per-class Precision, Recall and F1

RAG's effect on precision behaved like a class-dependent action. In many cases, precision stayed high or improved slightly along with recall, indicating that the additional hits retrieved by RAG were mostly correct. For example, GPT-4.1 nano's precision on "Abuse towards child" rose from 66.7% to 79.6% with RAG, even as it began capturing far more true child abuse cases (meaning, it did not sharply increase false positives for that class). Likewise, GPT-5 mini's precision for "Harm towards others" ticked up from 86.3% to 86.7%, while its recall climbed to 92.9%, yielding a $F_1$ of 0.897. However, there were a few trade-offs where RAG improved recall at the cost of precision. Gemini 2.5 Pro's performance on "Panic attack" is illustrative: RAG enabled it to catch about 30% more of the panic cases (recall 56.3% - 86.1%), but its precision on that class dropped from 90.7% to 63.5%. This suggests that the model started labeling more instances as "Panic attack" (correctly identifying most of the previously missed ones but also introducing some false positives). Yet, the $F_1$ for Gemini 2.5 Pro on panic queries improved substantially (from 0.694 to 0.731), indicating the recall gains outweighed the precision losses in net effect.

A similar pattern was seen with "Trauma" and "Prompt injection" for some models: those that were overtly conservative originally gained recall with RAG but sometimes misfired on a few easy cases, slightly reducing precision. Notably, majority "No risk" class predictions sometimes became less precise under RAG. Several models began flagging more inputs as risky (reducing false-negatives in minority classes), which inevitably meant a few truly safe inputs were misclassified as risks (reducing recall on "No risk" class). For instance, Gemini 2.5 Pro's recall fell by 14.9 points (90.1% - 75.2%) with RAG, and its "No risk" precision only rose slightly (78.7% - 81.5%), resulting in a significantly lower $F_1$ (0.7245 - 0.6364). In contrast, other models managed a better balance: o4-mini saw a modest drop in "No risk" recall (90.1% - 83.6%) but a sizable precision boost (59.0% - 69.5%), ultimately still improving its "No risk" $F_1$ (0.713 - 0.759). Overall, across all models and classes, RAG produced net gains in $F_1$ for 38 out of 42 class-level evaluations (7 intents across 6 models), while 4 cases showed slight declines.

### Inter-Model Agreement

Another notable outcome of RAG is its effect on agreement between models. Pairwise prediction agreement rates and Cohen's Kappa (κ) were computed between every pair of models, separately for the RAG and no-RAG conditions. Under no-RAG, the models often disagreed substantially, especially with varying accuracy counts. For example, GPT-4.1 nano's predictions matched Claude Sonnet 4 on only about 60.6% of instances without RAG. Its agreement with GPT-5 mini was even lower at ~49.0%, reflecting the tendency of a lightweight model like GPT-4.1 nano to misclassify many cases that the larger models got right. With RAG however, the models' outputs became more aligned. GPT-4.1 nano (with RAG) agreed with Claude on 75.1% of the test cases and with GPT-5 mini on ~72.98%. In fact, RAG increased the pairwise agreement between GPT-4.1 nano and every other model by ~15 - 25%.

| Model | Average Model Consensus |
|---|---|
| GPT-4.1 nano | 0.568 |
| o4-mini | 0.818 |
| Gemini 2.5 flash | 0.836 |
| Gemini 2.5 Pro | 0.758 |
| Claude Sonnet 4 | 0.839 |
| GPT-5 mini | 0.863 |

**Table 3:** Inter-model agreements

More generally, the spread in performance between models narrowed under RAG, leading to more consensus. Even models that were originally dissimilar showed markedly improved concordance. For instance, Gemini 2.5 Pro (the only model that didn't improve overall with RAG) still became more similar to others: its agreement with o4-mini rose from 71.8% to 74.94% and with Claude from 77.08% to 78.39%. The overall trend retained that RAG improved inter-model agreement, indicating that the models not only became more accurate individually, but also more consistent as a group in the labels they assigned.

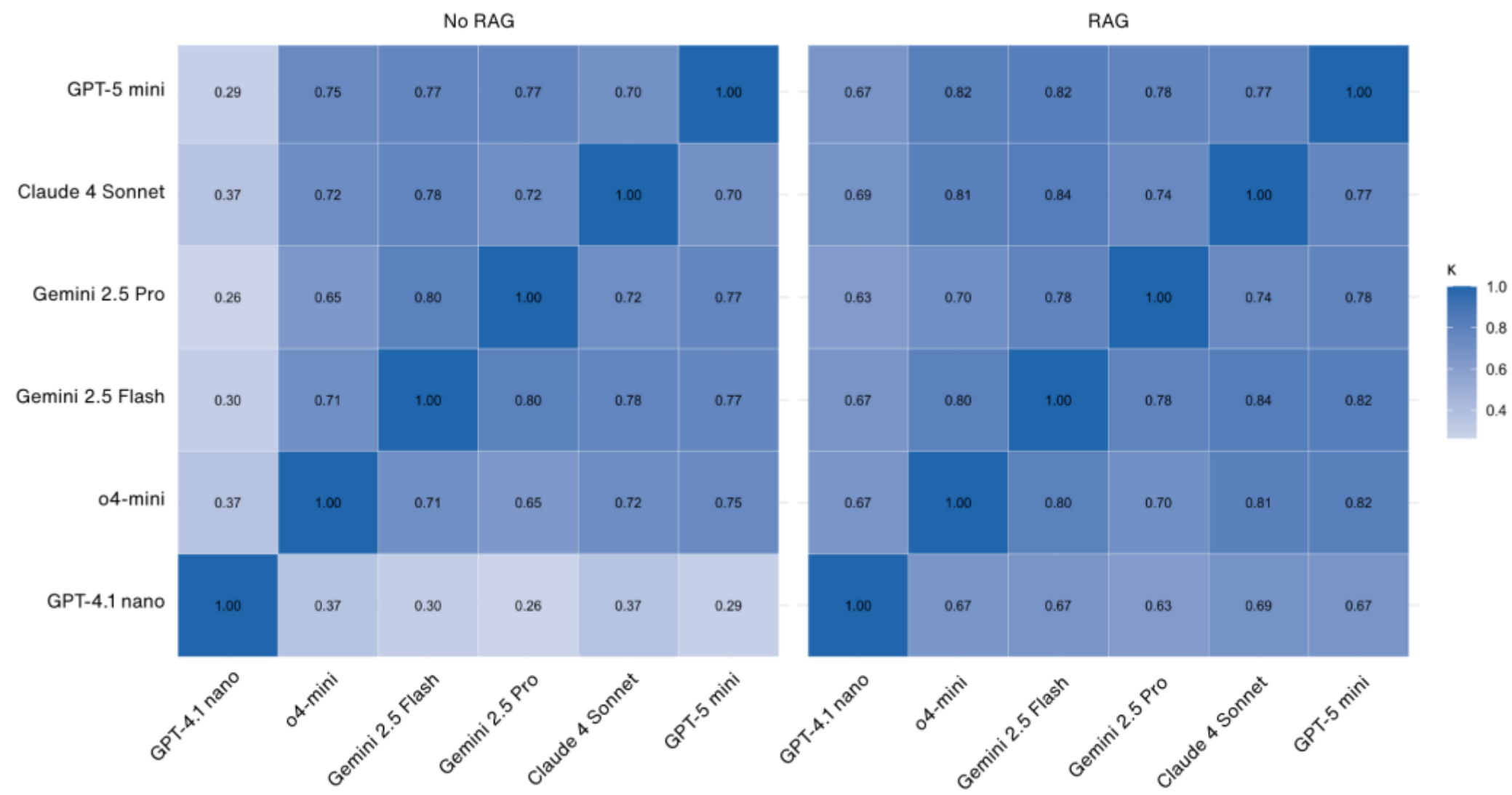

**Figure 4:** Cohen's κ Pairwise Agreements

Cohen's κ (multiclass) reveals the same pattern while accounting for chance agreement. Without RAG, many model pairs had only moderate κ values despite high raw agreement on the majority class. However, RAG enabled high κ values to approach the "almost perfect" agreement range, reinforcing that the ensemble of models reached a much higher consensus with RAG. In practical terms, this means that given the same query, the models' classifications were more likely to coincide when retrieval was available, presumably because the retrieved evidence guided them toward similar conclusions.

## Model Consistency

Intra-model consistency was examined for each model's outputs between the RAG and no-RAG conditions. This analysis essentially calculates the number of test instances where RAG changed its prediction. The answers varied widely by model capacity. GPT-5 mini was the most stable, allocating the same classification with and without RAG for ~88% of the test cases, altering its decision only ~12% of the time (201/1680 instances discordant). Claude Sonnet 4 was similarly consistent, with ~84.5% of outputs unchanged by RAG (discordant in 260 cases). In contrast, the smallest model GPT-4.1 nano changed its prediction on 39% of instances (659/1680) when using RAG, reflecting the earlier point that RAG heavily rewrote GPT-4.1 nano's decision profile. The mid-range models fell in between: o4-mini changed ~17% of the time ($b + c = 293$), and Gemini 2.5 flash ~14% (242 changes). Interestingly, Gemini 2.5 Pro's predictions shifted on about 21.5% of instances (361/1680) with RAG (more than one might expect given its overall accuracy didn't improve). A closer inspection found that many of Pro's changes with RAG were different errors rather than net improvements, e.g. it started catching some "Panic attacks" it previously missed but also began missing some "No risk" cases it previously got right, roughly cancelling out. In contrast, for GPT-4.1 nano and o4-mini the vast majority of changes were beneficial corrections ($c \gg b$, as noted). Overall, these consistency rates indicate that larger models are inherently more stable, they only adjust on a small subset of queries when additional context is provided, presumably those borderline cases where retrieval offers new clues. Smaller models, lacking robust internal knowledge, undergo more drastic output revision under RAG. From an ensemble perspective, RAG made the weaker models behave more like the stronger ones, hence the increased group agreement discussed above. We also note that even when models changed their answers under RAG, they usually changed from a wrong answer to the right one (for significant-improvement models), which is exactly the intended effect of retrieval augmentation.

## Error Analysis & Confusion Patterns

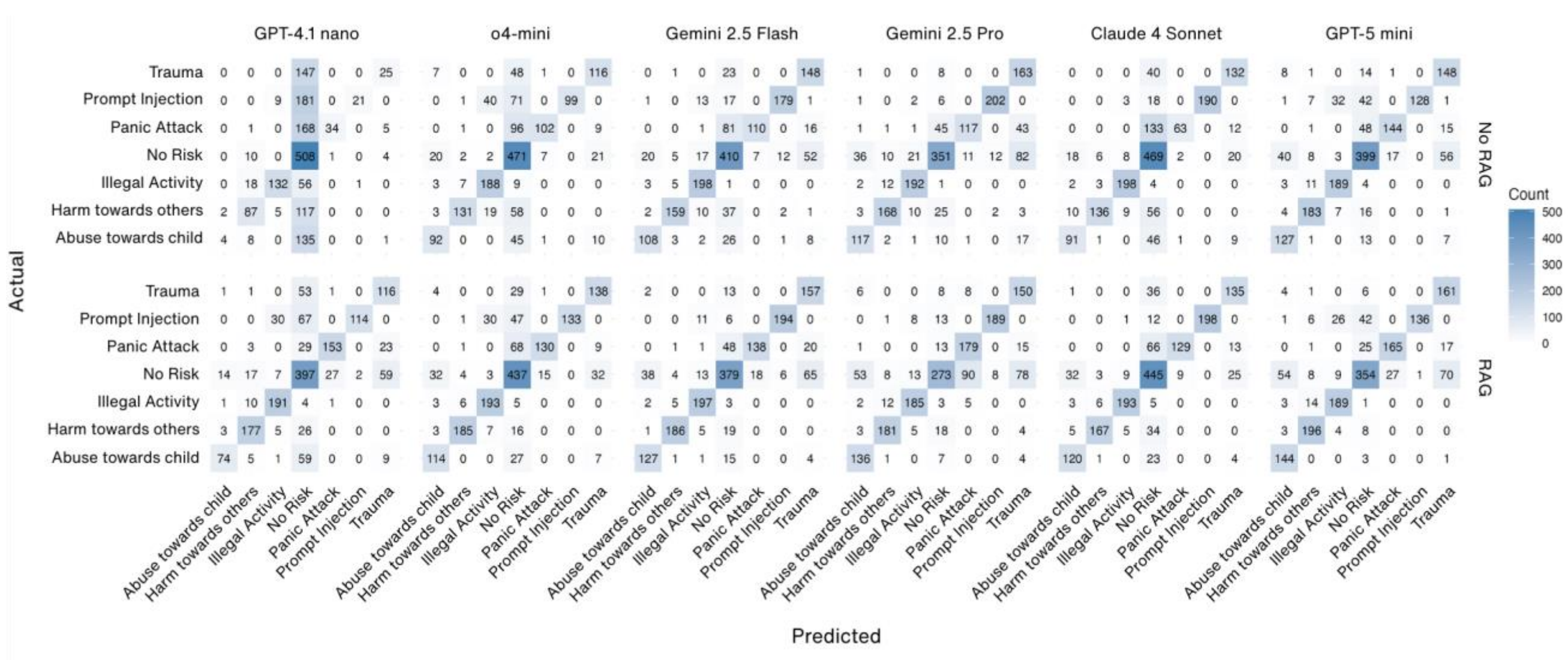

**Figure 5:** Confusion Matrices (Risk Category vs Model)

The class confusion matrices provide further insight into the specific error reductions achieved by RAG. Without RAG, a dominant error mode for multiple models was over-predicting the “No risk” class, i.e. labeling volatile queries as safe. For example, GPT-4.1 nano’s raw confusion matrix (i.e. no-RAG) shows that for actual “Abuse towards child” cases, the model overwhelmingly predicted “No risk” in almost all 148 instances (hence its 97% recall on “No risk” vs only 2.7% on child abuse). Similar patterns were seen for “Panic attack” and “Trauma”, which were frequently confused with “No risk” or other benign categories. After enabling RAG, these confusions diminished dramatically. The matrices show far more weight on the diagonal for those classes, e.g. GPT-4.1 nano correctly classifies 74 out of 148 child-abuse cases with RAG (50% recall) instead of only 4/148 without RAG. Likewise, Claude Sonnet 4, which initially misidentified “Panic attack” cases as “Harm towards others” or “No risk”, managed to correctly identify twice as many of them with RAG (raising recall from 30% to 62%). These improvements indicate that RAG supplied relevant cues or knowledge that helped disambiguate user intents that the models previously missed. Another prevalent confusion was between certain volatile categories themselves, e.g. distinguishing “Panic attack” vs “Trauma”, or “Abuse towards child” vs “Harm towards others”, which can be subtle. The RAG condition slightly alleviated some of these as well (e.g. models better separated trauma from general harm), though residual cross-confusions remained in some cases. On the other hand, new error patterns introduced by RAG were relatively few. One observable issue was an increase in false-positives for some volatile classes: e.g. Gemini 2.5 Pro with RAG sometimes misclassified harmless inputs as “Panic attack” or “Abuse”, contributing to its drop in “No risk” recall. These are cases where the retrieved content might have been misleading or the model over-interpreted a benign query as risky. Such errors highlight that RAG is not infallible, it can shift a model’s mistakes from “misses” to “false alarms.” However, from a safety perspective, many applications may prefer false positives (flagging a safe query for review) over false negatives (missing a truly risky query). Nonetheless, this is a trade-off to consider.

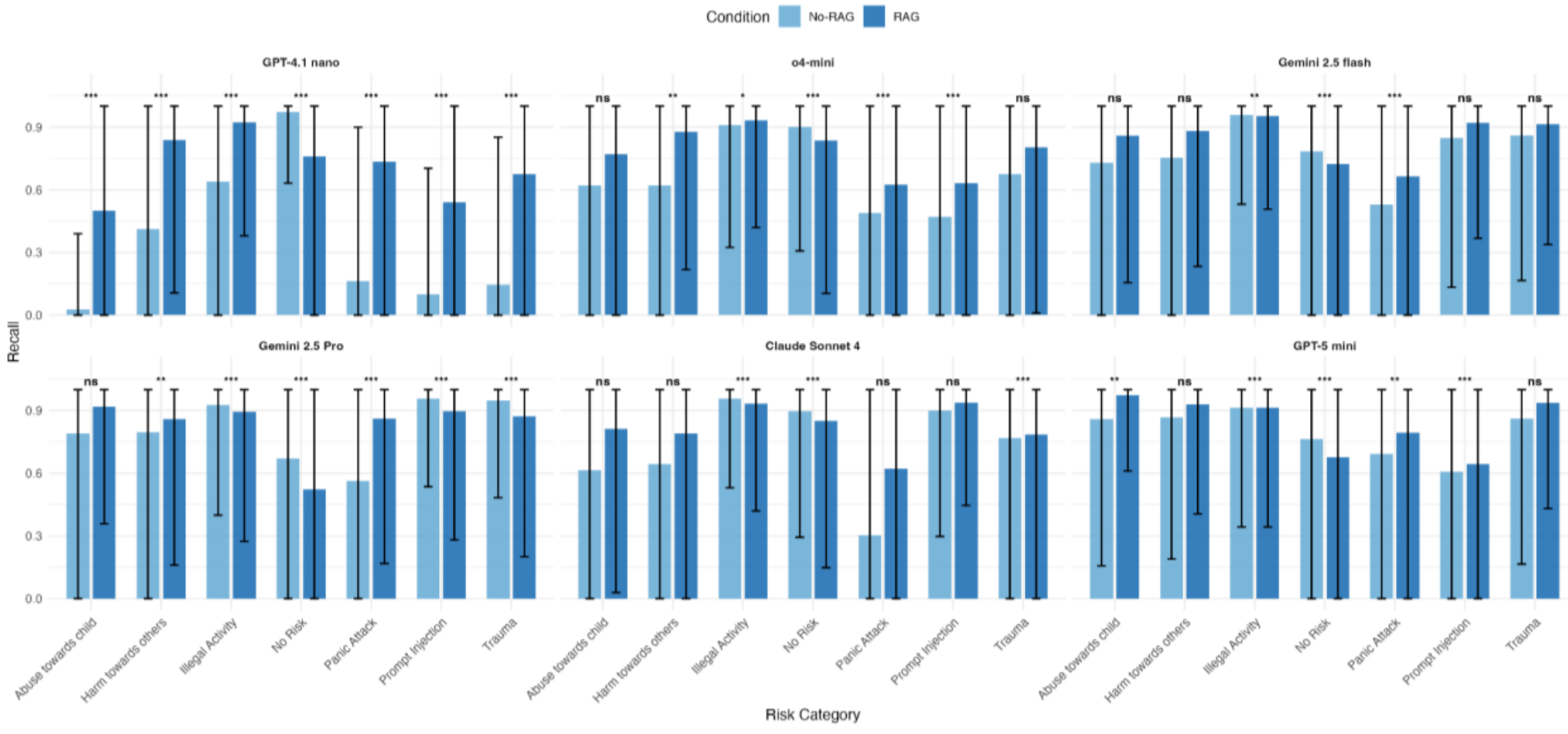

**Figure 6:** Per-class Recall with 95% Wilson CI and McNemar's Test Significance (RAG vs. No-RAG)

Where,
*****:** $p < 0.001$
****:** $p < 0.01$
***:** $p < 0.05$
**ns:** Not Significant

Statistical grounding for these observed shifts in error patterns is provided in Figure 6, which presents per-class recall with 95% Wilson confidence intervals and per-class McNemar's test results, confirming the significance of RAG-driven changes across risk categories and models.

Overall, the confusion matrices confirm that RAG's net impact was to reduce the most severe errors (failing to detect intent), even if it introduced a few mild confusions in return. The group consensus results support this: when all models' predictions disagree, it is often on those borderline cases where RAG's influence varied, but such instances became rarer with RAG. Under RAG, a greater fraction of queries were unanimously or majority-voted into the correct class by the model ensemble, whereas without RAG there were more cases where at least one model (often the weaker ones) strayed with an incorrect label.

## Discussion

### The Precision-Recall Tradeoff

In this evaluation, RAG-enabled models flagged fewer false negatives than their standalone counterparts in mental health contexts, particularly in the risk domains of child abuse, panic attacks and trauma. This result aligns with the findings of Lopez et al[14] and Xu et al[15], who observed that retrieval-augmented models in clinical settings significantly improved sensitivity (recall) and reduced missed cases compared to standalone LLMs.

Augmenting generative models with relevant knowledge retrieval makes them more sensitive to critical risk language, significantly reducing the risk of missing instances of genuine risk.[16] RAG shifts the balance towards higher recall of crises at the expense of precision as systems are explicitly designed to maximize the capture of all relevant contextual evidence (high recall), even when this means retrieving additional extraneous material and thereby reducing precision.[17]

Notably, per-class McNemar's tests revealed that these recall gains were statistically significant ($p < 0.001$) for most high-risk categories, though classes where models already performed strongly at baseline showed non-significant changes, suggesting that RAG's sensitivity-boosting effect is most pronounced where parametric knowledge is weakest. This asymmetry is particularly consequential in conversational mental health applications like Wysa, where false negatives and false positives carry fundamentally different clinical stakes. A flagged conversation triggers automated redirection to helplines and crisis resources within the system's escalation layer, which is a low-friction outcome. A missed detection, by contrast, means that a user in genuine crisis receives no escalation at all.[18]

#### Differential Impact with Model Capacity

Model capacities mediated the benefit gained from enabling RAG. RAG enabled models showed disproportionately large improvements in smaller LLMs, whereas larger models showed marginal change. GPT-4.1 nano showcased drastic performance improvements with classification accuracy and recall surging to thresholds equivalent to those of larger models. This was further confirmed at the class level, where GPT-4.1 nano was the only model to show statistically significant recall improvements across all seven risk categories ($p < 0.001$), while larger models showed selective significance concentrated in categories where their baseline recall was comparatively weaker. This is consistent with existing literature, facilitating the conclusion that RAG enabled systems can often rival or surpass larger models.[19] In this analysis, moderately sized retrieval was enough to compensate for limitations in safety domains or implicit intent detection.

By contrast, larger models derived little measurable benefit from RAG. This suggests a substantial value of pre-encoded knowledge or inference capabilities. For such models, retrieving additional information provided redundant context, yielding only minor refinements at best. Some larger models also exhibited an occasional override of correct internal judgement (without RAG), surfacing that retrieval may introduce minor noise. These observations echo the notion of diminishing returns for augmentation as model size grows.[20]

**Agreement and Consistency**

An additional effect of RAG was its influence on model agreement and consistency. It was observed that enabling retrieval led to a marked increase in inter-model agreement. This harmonization effect affirmed that retrieval supplied an external knowledge cue that all models could latch onto.[20]

We also examined intra-model consistency, where each model's own classification pattern changed with retrieval. The data suggested that smaller models became more consistent in flagging certain cues. However, for larger models, minimal change was noticed. This emphasized that RAG could shift a model's internal classification tendencies, especially for smaller models. It effectively re-calibrates models like GPT-4.1 nano to be more uniform and assertive in identifying critical issues. This proves useful as a system that responds the same way to the same stimuli (or conceptually similar stimuli) is more predictable and transparent.[21]

In essence, retrieval served not only as an on-demand infusion of expert knowledge but also a curated repository of vetted clinical protocols and validated response frameworks.,[22] which is particularly crucial in mental health applications where oftentimes such services act as the last barricade to volatile situations. Such a system could, for instance, retrieve the appropriate response script or risk assessment checklist whenever it suspects intent suggesting a crisis, ensuring its next action is grounded in vetted knowledge rather than just the vagaries of neural weights. The results provide empirical support that a retrieval strategy materially improves safety performance.

### Implications and Future Direction

It has been expected that large scale LLMs, by virtue of extensive training, will implicitly learn to predict intent around crisis-related content.[23] Methodologically, isolating individual architectural components within a deployed DMHI to quantify their marginal safety contribution constitutes a broader research programme spanning AI safety in crisis detection, optimization of AI-delivered interventions across clinical contexts, and adaptation of these systems for low-resource settings. This ablation framework offers a transferable approach for bridging controlled evidence and real-world clinical impact.

This study can offer guidance to system designers on how model size and architecture influence the returns from retrieval. If using a smaller language model (e.g. for on-device deployments or to minimize costs), results encourage integrating a robust retrieval component because the payoff can be significant. From a broader perspective, these findings underscore a paradigm shift for AI safety in healthcare: rather than treating safety as purely a function of model censorship or static fine-tuning, one can treat it as an information retrieval problem.

## Limitations:

This evaluation constitutes a single-layer ablation rather than an end-to-end system assessment. The production system incorporates additional independent safety layers like rule-based filtering, deterministic escalation, and human review, which were excluded here by design to isolate retrieval's contribution. Consequently, the absolute accuracy and recall figures reported reflect component-level behavior under isolation and should not be read as the detection performance of the deployed system. The evaluation was conducted in a controlled research setting on anonymized historical and re-synthesized data.

## Future directions

Moving forward, broader evaluation and targeted training will be crucial to further improve crisis intent detection. Recent benchmarks that probe LLM responses across diverse crisis scenarios reveal that even state-of-the-art models often miss subtly expressed distress cues and can produce inappropriate or formulaic responses.[24] These gaps underscore the need for better alignment and specialized tuning to handle nuanced expressions of crisis safely. One promising avenue is domain-specific fine-tuning of LLMs - for example, task-specific instruction tuning in the mental health domain has boosted performance so much that smaller specialized models matched or even outperformed much larger general models[25]. Another complementary direction is leveraging retrieval augmentation (RAG) to enhance smaller LLMs in safety-critical tasks. Injecting external knowledge via RAG can curb hallucinations and provide human-vetted context.[21] This aligns with observations that RAG-equipped systems can rival or surpass far bigger LLMs, yielding especially high returns at smaller scales. By combining efficient, lightweight models with robust retrieval of vetted crisis information (for example, well-tested intervention scripts and checklists), future systems could deliver cost-effective on-device crisis intent detection without sacrificing reliability.

## Conclusion

Ultimately, the study highlights that safety-aware behavior can be augmented through retrieval, and predicts that the next generation of digital mental health interventions will not only be empathetic and engaging, but also consistently safe, responsive to crises, and grounded in the same reliable knowledge and experience that human practitioners use. Such advancements bring us a step closer to AI systems that genuinely support clinicians and patients in tandem, combining the efficiency of automation with the prudence of evidence-based practice.

## Abbreviations

AI - Artificial Intelligence
LLM - Large Language Model
RAG - Retrieval Augmented Generation
PII - Personally Identifiable Information

## Acknowledgements
Not applicable.

## Funding
The study did not receive any funding.

## Author Information

### Authors and Affiliations
Wysa Inc., Boston, MA, United States
Anand Gupta, Akshat Surolia, Shubham Mishra, Shakil Imtiaz, & Chaitali Sinha

### Contributions
AG and CS conceived of and designed the study. AG, AK and SM contributed to the data collection of the study. SI: data analysis, drafting of the manuscript, revision and editing. CS: review and editing, supervision, project administration. All authors read and approved the final manuscript.

### Corresponding Author
Correspondence to Shakil Imtiaz.

## Ethics declarations

### Ethics approval and consent to participate
Not applicable.

### Consent for publication
Not applicable.

### Competing interests
AG, AS, SM, SI, and CS are employees of Wysa. The author(s) had access to anonymized user data for the purpose of this study.